\documentclass[twocolumn]{raa}   
\usepackage{graphicx,times}
\usepackage{natbib}
\usepackage{url,float,booktabs,array,calc}
\usepackage{hyperref}
\usepackage{xurl}
\usepackage{microtype}
\usepackage{caption}
\usepackage{stfloats} 

\begin{document}

\title{SVOM Real-time Response and Collaboration System}

\volnopage{Vol.0 (202x) No.0, 000--000}   
\setcounter{page}{1}                     

\author{Bai Meng\inst{1,2} \and Li Boquan\inst{1} \and Su Ju\inst{1}
        \and Feng Zhun\inst{1} \and Man Yichuan\inst{1} \and Xiao Zhigang\inst{1}\and Zhang Meng\inst{1}
        \and Liu Yurong\inst{1} \and Hu Tai\inst{1,\star}}

\institute{National Space Science Center, Chinese Academy of Sciences,
           Beijing 100190, China\\ \and
           University of Chinese Academy of Sciences, Beijing 100190, China\\
           \vs\no
           {\small Received 202x month day; accepted 202x month day}
}

\abstract{%
The SVOM mission (Space-based multi-band astronomical Variable Objects Monitor)
is a Franco-Chinese mission dedicated to the study of the most distant explosions
of stars, the gamma-ray bursts. Here, we introduce the real-time response and
collaboration system of SVOM, with the adoption of the BeiDou-3 short message
communication service. We present the SVOM on-board and on-ground system  \  {designs}
and data flow, together with the collaboration mechanism with other missions.
In the first year of the in-flight operation, SVOM has detected 172 gamma-ray
bursts, including 147 by the GRM instrument and 62 by the ECLAIRs instrument.
 \  {At the same time}, SVOM has performed 1040 observations, including 122  \  {ToO-EX(Target of Opportunity-Exceptional) observations,
48 ToO-MM(Target of Opportunity-Multi-messenger) observations and 870 ToO-NOM(Target of Opportunity-Nominal) observations.} All these have increased the  \  {scientific output of the mission.}
\keywords{real-time response --- BeiDou-3 short message communication --- collaboration mechanism}
}

\maketitle
\section{Introduction}\label{sec:intro}
The SVOM mission (Space-based multi-band astronomical Variable Objects Monitor) is a Franco-Chinese mission dedicated to the study of the most distant explosions of stars, the gamma-ray bursts. It was launched on June 22, 2024 by the Chinese Long March 2C rocket from the Xichang launch base. It is the result of a collaboration between the two national space agencies, CNSA (China National Space Administration) and CNES (Centre national d'études spatiales).

The mission consists of 4 main instruments of which 2 are French (ECLAIRs and MXT) and 2 are Chinese  \  {{(GRM and VT),}whose goals are described below:

\begin{itemize}
  \item The ECLAIRs telescope to detect and localize gamma bursts in the X-ray band and low-energy gamma rays (from 4 to 250 keV) (\citep{Godet+etal+2026}).
  \item The MXT telescope (Microchannel X-ray Telescope) for the observation of \  { gamma-ray burst afterglows} in the soft X-ray range (0.2 to 10 keV) \citep{Goetz+etal+2026}.
  \item The GRM (Gamma Ray Burst Monitor) to measure the spectrum of high-energy bursts (from 15 keV to 5000 keV) \citep{Sun+etal+2026}.
  \item The VT telescope (Visible Telescope) operating in the visible range to detect and observe the visible emission produced immediately after a gamma burst \citep{Qiu+etal+2026}.
\end{itemize}

While the main goal of the SVOM mission is to ensure the observation of about 100 gamma-ray bursts per year, it is also a formidable tool for probing the transient sky. SVOM is able to generate an alert after the detection of a transient phenomenon, thanks in particular to its large field of view instruments ECLAIRs and GRM. In parallel, SVOM is also reacting to  \  {targets of opportunity (ToO) alerts issued }from other transient sky observatories, on the ground or in space, and then points its instruments towards the object. At system level, the ToO system was designed to perform observations within 48 h for a standard target of opportunity (ToO-NOM), and within 12 h for an exceptional target opportunity (ToO-EX)(e.g. galactic supernova or GW alert). Multi-messenger ToOs (ToO-MM) were designed to search for counterparts of poorly localized events, such as gravitational-wave or high-energy neutrino alerts. For details on the SVOM mission see \cite{Cordier+etal+2026b}.

In case of a GRB trigger, the ECLAIRs, GRM and other important information will be downlinked as trigger alert messages to the ground.

In order to carry out rapidly follow-up observations, a real-time downlink and uplink loop is required. Considering the current status of the real-time downlink and uplink resources in China, SVOM  \  {adopted} the global short message communication service of BeiDou-3 navigation satellite system to downlink the trigger alert messages to the ground and uplink the follow-up observation to the satellite. SVOM is the first collaborative scientific satellite to use the BeiDou-3 global short message service on board and capable of real-time downlink and uplink \citep{yang2020bds}.

The SVOM Chinese Mission Science Ground Segment includes a dedicated component responsible for receiving, processing, and transmitting BeiDou short messages. In the following, we describe the onboard design and data flow in Section 2 and the ground real-time response and collaboration system in Section 3. We report the in-flight performance of the first year in Section 4. Finally, in Section 5 we give a summary.

\section{Onboard Design and Data Flow}\label{sec:onboard}
\subsection{Satellite Design of Trigger and BeiDou Communication}

 \  {The BeiDou} communication subsystem is a key asset for the SVOM satellite system to enhance the potential scientific outcome of the mission.  \  {For this scientific goal}, the main requirements of BeiDou short message system are divided into two main parts: upload ToO-EX and ToO-MM observations and download alert message to the ground.

 \  {The BeiDou} short message can be used to upload ToO-EX and ToO-MM  \  {observation requests}, including the ToO observation schedule and  \  {payload} configuration. The uplink frequency of BeiDou short message is 3 times/min.

For the downlink, it includes the trigger messages from the payload and critical housekeeping data from the satellite. The downlink alert messages include the ECLAIRs alert messages, GRM general triggers and light curve, spectrum messages, MXT positions, PDPU-GRB alert messages, VT attitude and simplified chart messages. The downlink frequency of BeiDou short message is 3 times/min. \  {The size of one BeiDou packet is about 70 bytes.\  {The simplified chart for VT refers to the downlink message transmitted via BeiDou, which is different from the normal chart for X-band downlink.}}

The transmission time delay of the BeiDou short message between Mission Center \citep{Liu+etal+2026} and satellite is less than 100s in 95\% cases. The availability of the BeiDou short message uplink and downlink (including BeiDou ground station and BeiDou \  {MEO(Medium Earth Orbit) network)} to the SVOM satellite is estimated at least 90\% for uplink and downlink.

\subsection{Board to Ground Data Flow}

There are \  {two} processes for board and ground, including the board to ground message transmission process and the ground to board message transmission process. The \  {two} processes are shown in Figure~\ref{fig1}.

When the SVOM instruments detect a transient source trigger message that must be transmitted to the ground immediately. The BeiDou global short message link is a feasible and rapid transmission channel. The SVOM satellite will select a BeiDou MEO(Medium Earth Orbit) satellite by signal strength and send the trigger message. This BeiDou MEO satellite will then forward the triggering message via the inter-satellite link to the BeiDou center. Finally, the BeiDou Center transmits the triggering information to the Mission Center via the BeiDou regional message link.

\  {When scientists identify a new transient source in outburst, they expect that the SVOM satellite will start observing as soon as possible.} The science center submits the observation requests to the Mission Center, which responds in real time, processes and generates the observation control instructions, and transmits them via the regional message link to the BeiDou center. The BeiDou center then forwards the mission instructions in real time via the global message link to the SVOM satellite. After receiving the observation instructions, \  {the SVOM satellite performs a rapid slew, and observes the source in outburst. }This process of ground-board message transmitting \  {can be as fast as few minutes.}

\begin{figure}
\centering
\includegraphics[width=\columnwidth]{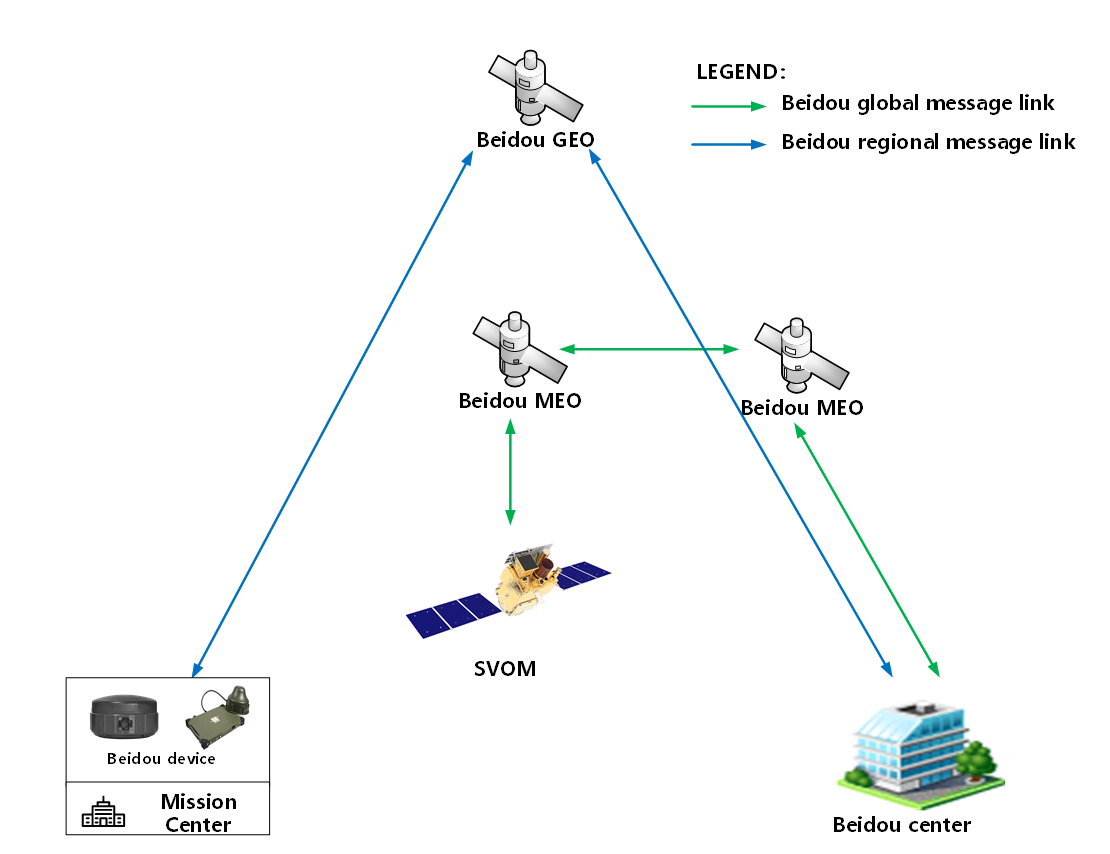}
\centering
\caption{Transmission process}
\label{fig1}
\end{figure}

\section{Ground Real-time Response and Collaboration System}\label{sec:ground}
\subsection{BeiDou Communication Design on Ground}

For the BeiDou communication design on ground, the BeiDou data interaction system includes the BeiDou device, data reception, data transition, data fusion and system monitoring.

\begin{figure}
\centering
\includegraphics[width=\columnwidth]{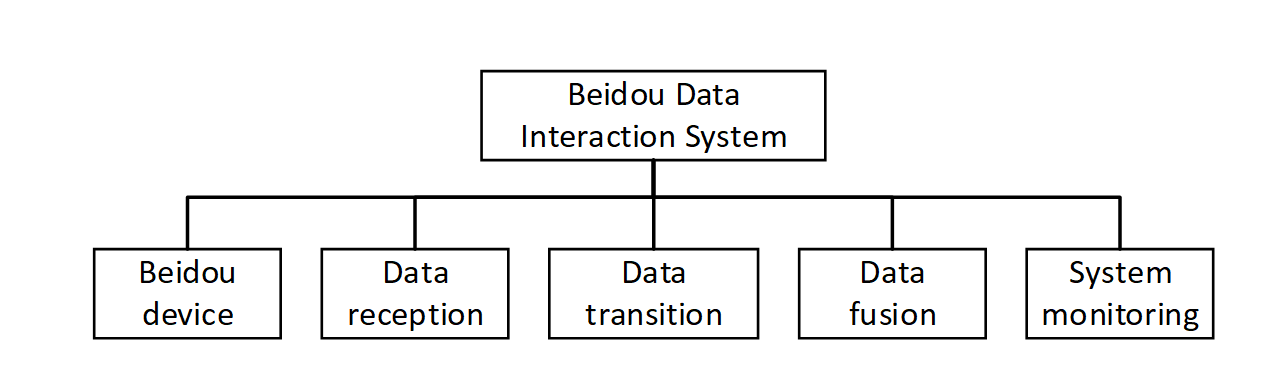}
\caption{Main components and function}
\label{fig2}
\end{figure}

The main functions are shown in Figure~\ref{fig2} and are as follows:

\begin{enumerate}
  \item \textbf{BeiDou device:} Carry out two-way data communication with the BeiDou satellites.
  \item \textbf{Data reception:} Obtain the data according to the interface protocol of the BeiDou device, including the trigger information of the SVOM satellite, the device status information, etc.
  \item \textbf{Data transition:} Receive the SVOM satellite observation control instructions, manage the message queue, and send them to the BeiDou device according to the interface protocol and sending strategy. Additionally, feedback
on the sending status.
  \item \textbf{Data fusion:} The data received from the primary and backup terminals will be merged and duplicated. Then forwarded to the Mission Center according to the interface protocol.
  \item \textbf{System monitoring:} Receive BeiDou device status information, determine the availability of the BeiDou communication link, monitor and alarm the link status.
\end{enumerate}

For the reliability design, it includes the following methods:

\begin{enumerate}
  \item \textbf{BeiDou device \  {warm backup:}} Using dual-beam backup of BeiDou devices, both have the capabilities of short message reception and transmission. The main and backup terminals
simultaneously receive trigger information from the SVOM satellite.
  \item \textbf{Data processing server \  {warm }backup:} The data processing server adopts dual-machine warm standby, with separate deployments of data processing software. The software exchange heartbeat information with each other and simultaneously receive the data. By default, the output results are generated by the software on the backup host machine. When the software on the host machine fails, the backup automatically switches to the host and outputs the results.
\end{enumerate}

\subsection{Quick Response System Design on Ground}

The quick response system design on ground is described in Figure~\ref{Figure3-20260205}.

\begin{figure}
\centering
\includegraphics[width=\columnwidth]{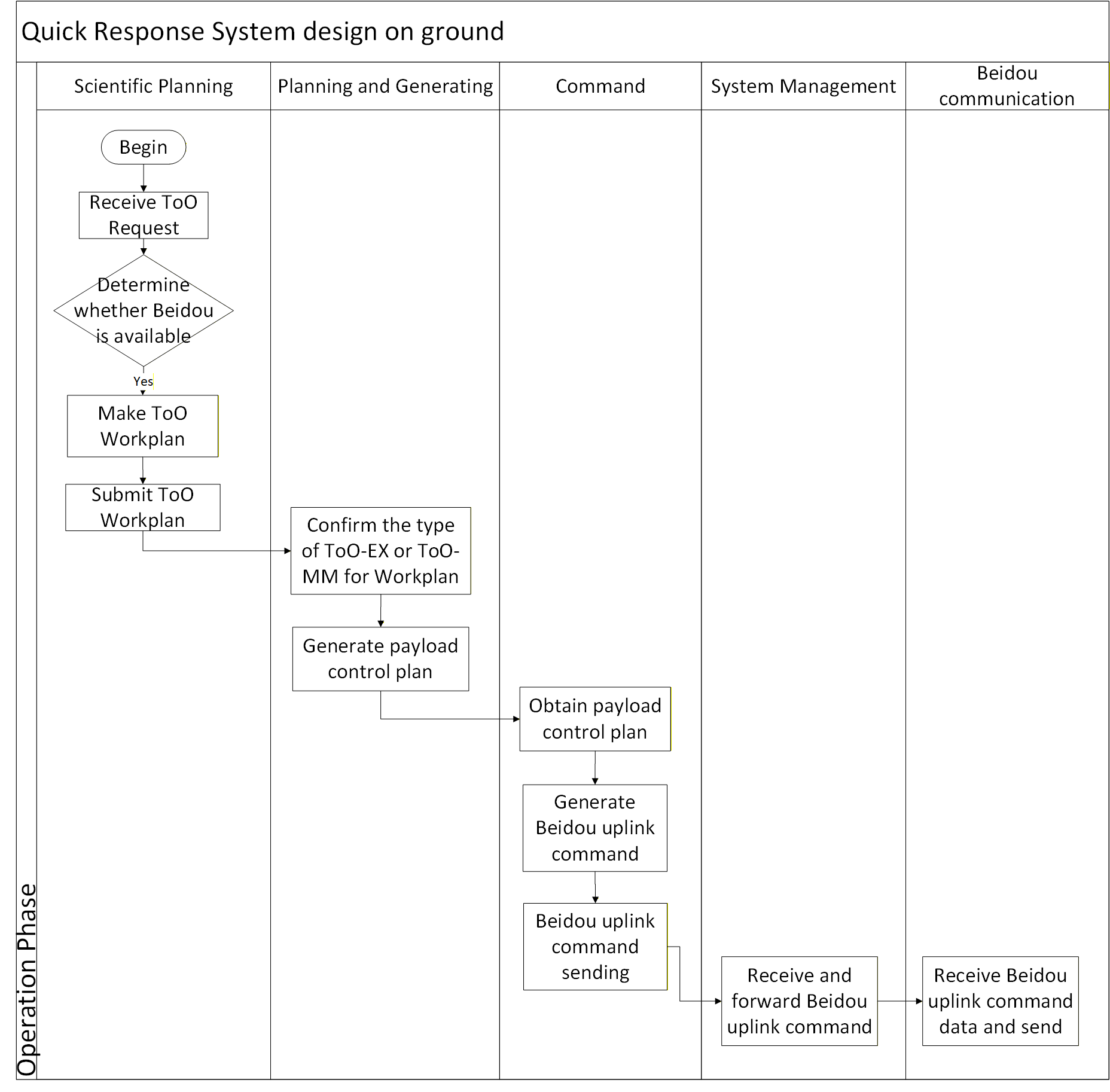}
\caption{The quick response system design on ground process}
\label{Figure3-20260205}
\end{figure}

\begin{enumerate}
  \item[a)] Chinese Science Centre (CSC) sends a ToO Request to Mission Center, and SVOM scientific planning service software develops a ToO WorkPlan based on the ToO observation plan. It determines the uplink mode of ToO based on whether BeiDou uplink is available and submits the ToO WorkPlan to SVOM planning software;
  \item[b)] SVOM planning software is used for planning and generating payload control plans;
  \item[c)] SVOM instruction generation and control software obtains the payload control plan, verifies it, and generates BeiDou uplink instructions. The instruction control function obtains BeiDou uplink instructions for instruction transmission;
  \item[d)] The BeiDou uplink command of the BeiDou interactive system is sent through the BeiDou front-end communication of the business communication and management scheduling software.
\end{enumerate}

\subsection{Collaboration Mechanism with Other Missions}

In this study, the joint observation and collaboration mechanism for astronomical satellites has been successfully established and put into operation. Centered on the SVOM mission, this mechanism collaborates with the Swift satellite \citep{Gehrels+etal+2004} and the Einstein Probe (EP) \citep{Yuan+etal+2025} satellite to build an efficient multi-satellite joint response network. When the ECLAIRs \  {telescope} aboard the SVOM satellite detects transient sources such as gamma-ray bursts, the alert information generated is transmitted in real time to French and Chinese scientific centres (\cite{Huang+etal+2026}, \cite{Louvin+etal+2026}) via the VHF network \citep{cordier+etal+2026a}, supplemented by the BeiDou system.

Once this alert information has been processed, if the GRB detection is validated, a ToO request is send to the Swift Satellite Science Operation Center and the EP Satellite Operation Center through an automated process.

Thus, upon receiving the alert, each science operations center immediately initiates a rapid response process and formulates corresponding Target of Opportunity (ToO) observation plans based on the alert information. These observation commands are uploaded to the respective satellites with minimal delay through a combination of BeiDou short-message service and normal telemetry \& command channels. Through this collaborative mechanism, the three satellites can complete observation attitude adjustment and payload configuration in the shortest possible time, enabling multi-wavelength and\  { multi-telescopes} joint observation of the same target source, which significantly improves the efficiency of time-domain astronomical observation and the value of scientific output.

This innovative joint observation model not only demonstrates the collaborative advantages of international cooperation but also provides a new research paradigm for understanding extreme physical processes in the universe through multi-messenger and multi-platform data fusion.

\section{In-flight Performance}\label{sec:performance}
\subsection{SVOM In-flight Performance of ToO Observation}

\begin{figure}
\centering
\includegraphics[width=\columnwidth]{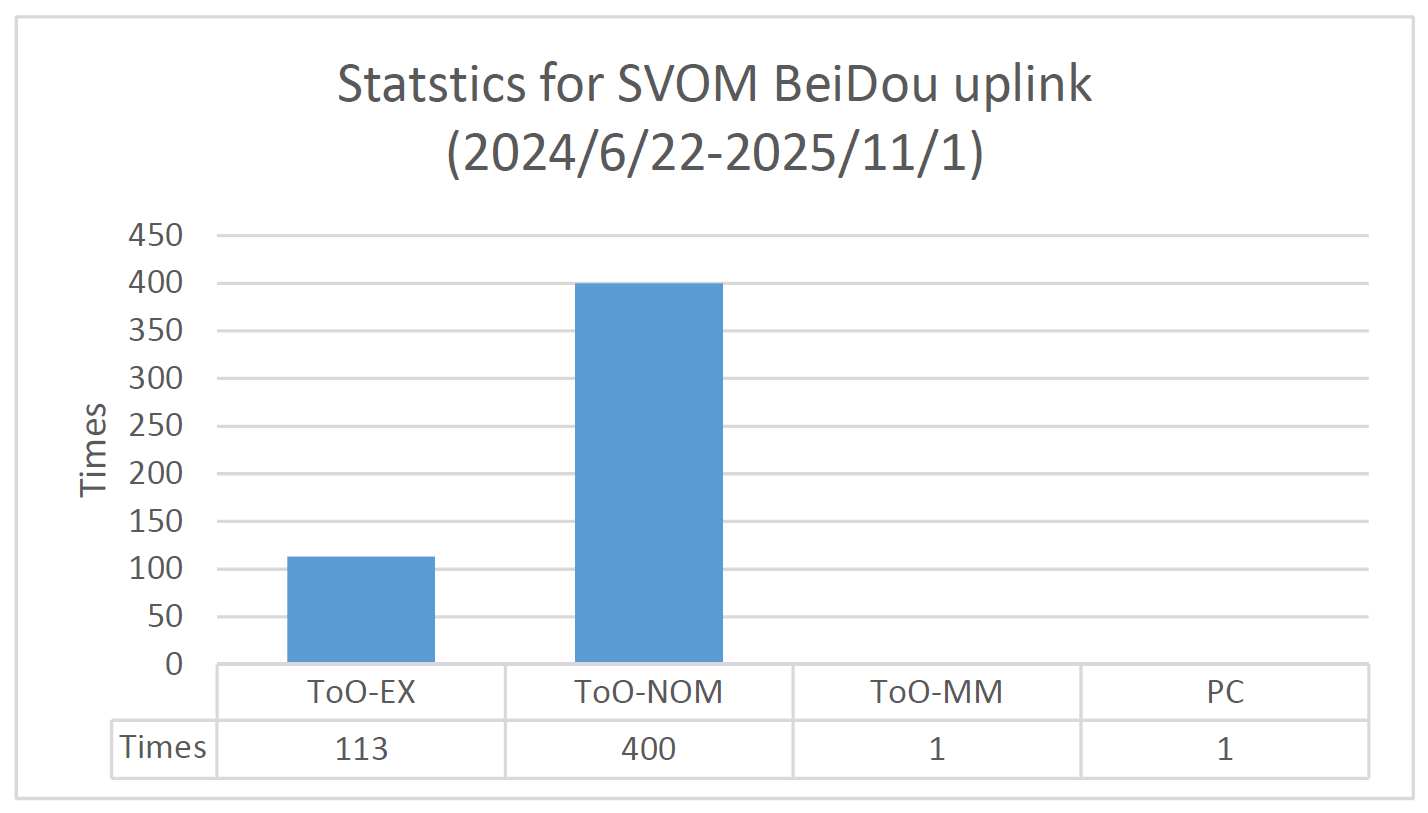}
\caption{ToO statistics}
\label{Figure4-20260206}
\end{figure}

From satellite launch to November 1, 2025, SVOM satellite has uploaded 113 ToO-EX plans, 400 ToO-NOM plans, 1 ToO-MM plan through BeiDou, and urgently\  { authorized} one PC(Payload Configuration) plan, see Figure~\ref{Figure4-20260206} .

The average duration from receiving the request to the complete successful uplink of the instruction is less than 30 minutes; the minimum duration is less than 2 minutes.

\begin{table}
\centering
\caption{Time statistics for ToO uplink}
\resizebox{0.95\linewidth}{!}{
\begin{tabular}{@{}lcc@{}}
\toprule
\textbf{Type} & \textbf{Average Duration (min)} & \textbf{Minimum Duration (min)} \\
\midrule
ToO-EX  & 20 & 2 \\
ToO-NOM & 24 & 2 \\
\bottomrule
\end{tabular}
}
\end{table}

\subsection{Collaborative Observation of SVOM with EP and Swift}

The collaborative observation performance is presented in this section. 

For the joint observation with SVOM and Swift, there are already \  {46 numbers of observation in 2025 on the date of Nov.30}. The statistics is shown in the Table2 (see end of paper), \  {with the newly discovered unknown transient sources (uncatalogued sources) and all detected radiation sources in the batch (total sources), enabling astronomers to quickly identify new targets. This is also the core objective of the joint observation by the  \  {two }satellites: to capture uncatalogued transient sources such as gamma-ray bursts.}

\begin{table*}[b] 
\centering
\normalsize
\caption{Joint SVOM/Swift observations in 2025 .}
\label{tab:swift-main}
\setlength{\tabcolsep}{25pt}
\begin{tabular}{@{}rllc@{}}
\toprule
No & Name & Uncatalogued sources(Total sources) & Trigger Time(UTC)\\

\midrule
1  & \href{https://www.swift.ac.uk/SVOM/SVOM_FIELD00047/}{GRB\,251103A} & 1\,(2)  & 2025-11-03 04:46:26\\
2  & \href{https://www.swift.ac.uk/SVOM/SVOM_FIELD00046/}{GRB\,251027A} & 3\,(0)  & 2025-10-27 09:37:56\\
3  & \href{https://www.swift.ac.uk/SVOM/SVOM_FIELD00045/}{GRB\,251026A} & 2\,(0)  & 2025-10-26 08:26:46\\
4  & \href{https://www.swift.ac.uk/SVOM/SVOM_FIELD00044/}{GRB\,251025B} & 1\,(0)  & 2025-10-25 14:24:00\\
5  & \href{https://www.swift.ac.uk/SVOM/SVOM_FIELD00043/}{sb25102501}  & 0\,(1)  & 2025-10-25 09:32:38\\
6  & \href{https://www.swift.ac.uk/SVOM/SVOM_FIELD00042/}{GRB\,251016A} & 32\,(1) & 2025-10-16 14:59:19\\
7  & \href{https://www.swift.ac.uk/SVOM/SVOM_FIELD00041/}{GRB\,251013C} & 1\,(1)  & 2025-10-13 17:39:42\\
8 & \href{https://www.swift.ac.uk/SVOM/SVOM_FIELD00040/}{sb25101004}  & 4\,(3)  & 2025-10-10 03:47:53\\
9 & \href{https://www.swift.ac.uk/SVOM/SVOM_FIELD00039/}{GRB\,251002A} & 3\,(0)  & 2025-10-02 20:14:49\\
10 & \href{https://www.swift.ac.uk/SVOM/SVOM_FIELD00038/}{sb25091517}  & 3\,(2)  & 2025-09-15 21:56:13\\
11 & \href{https://www.swift.ac.uk/SVOM/SVOM_FIELD00037/}{sb25091206}  & 1\,(0)  & 2025-09-12 21:06:23\\
12 & \href{https://www.swift.ac.uk/SVOM/SVOM_FIELD00036/}{sb25091001}  & 1\,(0)  & 2025-09-10 03:16:10\\
13 & \href{https://www.swift.ac.uk/SVOM/SVOM_FIELD00035/}{sb25090605}  & 0\,(2)  & 2025-09-06 21:18:25\\
14 & \href{https://www.swift.ac.uk/SVOM/SVOM_FIELD00034/}{GRB\,250903A} & 4\,(0)  & 2025-09-03 17:24:43\\
15 & \href{https://www.swift.ac.uk/SVOM/SVOM_FIELD00033/}{GRB\,250901A} & 1\,(1)  & 2025-09-01 18:14:17\\
16 & \href{https://www.swift.ac.uk/SVOM/SVOM_FIELD00032/}{GRB\,250831A} & 3\,(1)  & 2025-08-31 09:28:52\\
17 & \href{https://www.swift.ac.uk/SVOM/SVOM_FIELD00031/}{GRB\,250818B} & 1\,(0)  & 2025-08-18 03:29:08\\
18 & \href{https://www.swift.ac.uk/SVOM/SVOM_FIELD00030/}{sb25081602}  & 0\,(2)  & 2025-08-16 15:31:10\\
19 & \href{https://www.swift.ac.uk/SVOM/SVOM_FIELD00029/}{sb25081303}  & 4\,(0)  & 2025-08-13 22:51:15\\
20 & \href{https://www.swift.ac.uk/SVOM/SVOM_FIELD00028/}{GRB\,250812A} & 1\,(0)  & 2025-08-12 02:45:02\\
21 & \href{https://www.swift.ac.uk/SVOM/SVOM_FIELD00027/}{GRB\,250808A} & 1\,(0)  & 2025-08-08 18:23:46\\
22 & \href{https://www.swift.ac.uk/SVOM/SVOM_FIELD00026/}{GRB\,250806A} & 1\,(0)  & 2025-08-06 07:57:54\\
23 & \href{https://www.swift.ac.uk/SVOM/SVOM_FIELD00025/}{GRB\,250727A} & 2\,(0)  & 2025-07-27 10:53:39\\
24 & \href{https://www.swift.ac.uk/SVOM/SVOM_FIELD00024/}{GRB\,250713A} & 2\,(0)  & 2025-07-13 17:06:02\\
25 & \href{https://www.swift.ac.uk/SVOM/SVOM_FIELD00023/}{GRB\,250706B} & 1\,(0)  & 2025-07-06 17:05:43\\
26 & \href{https://www.swift.ac.uk/SVOM/SVOM_FIELD00022/}{GRB\,250704A} & 1\,(2)  & 2025-07-04 03:42:20\\
27 & \href{https://www.swift.ac.uk/SVOM/SVOM_FIELD00021/}{sb25062804}  & 1\,(1)  & 2025-06-28 16:29:10\\
28 & \href{https://www.swift.ac.uk/SVOM/SVOM_FIELD00020/}{sb25061218}  & 2\,(0)  & 2025-06-12 20:34:38\\
29 & \href{https://www.swift.ac.uk/SVOM/SVOM_FIELD00019/}{sb25061207}  & 4\,(0)  & 2025-06-12 10:27:02\\
30 & \href{https://www.swift.ac.uk/SVOM/SVOM_FIELD00018/}{GRB\,250610B} & 4\,(3)  & 2025-06-10 16:33:19\\
31 & \href{https://www.swift.ac.uk/SVOM/SVOM_FIELD00017/}{GRB\,250530A} & 4\,(0)  & 2025-05-30 06:31:54\\
32 & \href{https://www.swift.ac.uk/SVOM/SVOM_FIELD00016/}{sb25051002}  & 3\,(0)  & 2025-05-10 04:38:42\\
33 & \href{https://www.swift.ac.uk/SVOM/SVOM_FIELD00015/}{GRB\,250507A} & 3\,(0)  & 2025-05-07 06:34:52\\
34 & \href{https://www.swift.ac.uk/SVOM/SVOM_FIELD00015/}{GRB\,250506A} & 1\,(0)  & 2025-05-06 02:23:22\\
35 & \href{https://www.swift.ac.uk/SVOM/SVOM_FIELD00013/}{sb25042207}  & 3\,(1)  & 2025-04-22 22:36:17\\
36 & \href{https://www.swift.ac.uk/SVOM/SVOM_FIELD00012/}{GRB\,250419A} & 2\,(1)  & 2025-04-19 02:29:32\\
37 & \href{https://www.swift.ac.uk/SVOM/SVOM_FIELD00011/}{sb25041603}  & 1\,(5)  & 2025-04-16 06:28:04\\
38 & \href{https://www.swift.ac.uk/SVOM/SVOM_FIELD00010/}{GRB\,250403A} & 16\,(0)& 2025-04-03 15:15:20\\
39 & \href{https://www.swift.ac.uk/SVOM/SVOM_FIELD00009/}{GRB\,250402A} & 3\,(0)  & 2025-04-02 23:08:01\\
40 & \href{https://www.swift.ac.uk/SVOM/SVOM_FIELD00008/}{sb25032902}  & 1\,(0)  & 2025-03-29 12:10:18\\
41 & \href{https://www.swift.ac.uk/SVOM/SVOM_FIELD00007/}{sb25032901}  & 2\,(0)  & 2025-03-29 04:17:52\\
42 & \href{https://www.swift.ac.uk/SVOM/SVOM_FIELD00006/}{sb25032803}  & 2\,(0)  & 2025-03-28 16:56:56\\
43 & \href{https://www.swift.ac.uk/SVOM/SVOM_FIELD00005/}{sb25032706}  & 3\,(1)  & 2025-03-27 21:11:43\\
44 & \href{https://www.swift.ac.uk/SVOM/SVOM_FIELD00004/}{sb25032302}  & 3\,(3)  & 2025-03-23 04:08:21\\
45 & \href{https://www.swift.ac.uk/SVOM/SVOM_FIELD00003/}{GRB\,250317B} & 2\,(0)  & 2025-03-17 02:12:18\\
46 & \href{https://www.swift.ac.uk/SVOM/SVOM_FIELD00002/}{GRB\,250314A} & 3\,(0)  & 2025-03-14 12:56:42\\
47  & \href{https://www.swift.ac.uk/SVOM/SVOM_FIELD00001/}{sb25021202}  & 0\,(0)  & 2025-02-12 15:02:40\\

\bottomrule
\end{tabular}
\end{table*}

\  {Concerning the joint observations of SVOM and EP, there are already 34 observations in 2025 as of November 30. All the triggers, the generation of ToO workplans, and uplinks are automatic.}

\section{Conclusions and Perspectives}\label{sec:concl}
SVOM is a Franco-Chinese mission dedicated to the study of the most distant explosions of stars, the gamma-ray bursts, using the BeiDou-3 short message service. During the first year of operation, SVOM \  {has triggered} 184 gamma-ray bursts.

The real-time response and collaboration system based on the BeiDou short message service of SVOM establishes a low-latency and highly reliable satellite-ground cooperative link by deeply integrating real-time short message communication with traditional TT\&C (Tracking, Telemetry, and Command) channels.

This design enables\  { second-level(refers to a time scale measured in seconds)} transmission of transient source alerts from the satellite to the ground center, supports parallel delivery and rapid uplink of multi-satellite observation commands, significantly shortens the joint response cycle of multiple satellites, and greatly improves the efficiency of on-orbit satellite attitude adjustment and payload configuration.

\noindent\textbf{Acknowledgments} \
This work is supported by the Strategic Priority Research Program on Space Science,
the Chinese Academy of Sciences (grant Nos.\ XDA15040100 and XDA15040400).

\bibliographystyle{raa}   
\bibliography{bibtex}   

\end{document}